\documentclass[preprint,aps,floats,nofootinbib]{revtex4}

\usepackage{graphicx}
\usepackage{amsmath}
\usepackage{subfigure}

\newcommand{\btopik}{B \to \pi K}

\newcommand{\beq}{\begin{equation}}
\newcommand{\eeq}{\end{equation}}
\newcommand{\bea}{\begin{eqnarray}}
\newcommand{\eea}{\end{eqnarray}}
\newcommand{\ba}{\begin{array}}
\newcommand{\ea}{\end{array}}
\newcommand{\bi}{\begin{itemize}}
\newcommand{\ei}{\end{itemize}}
\newcommand{\bn}{\begin{enumerate}}
\newcommand{\en}{\end{enumerate}}
\newcommand{\bc}{\begin{center}}
\newcommand{\ec}{\end{center}}
\renewcommand{\l}{\left}
\renewcommand{\r}{\right}

\renewcommand{\ol}{\overline}
\newcommand{\De}{\Delta}
\newcommand{\al}{\alpha}

\newcommand{\ga}{\gamma}
\newcommand{\de}{\delta}
\newcommand{\la}{\lambda}

\newcommand{\nl}{\nonumber\\}

\newcommand{\wt}[1]{\widetilde{#1}}

\newcommand{\BsBs}{B_s^0-\overline{B}_s^0}
\newcommand{\btos}{\bar{b} \to \bar{s}}
\begin{document}

\vspace*{0.5cm}

\title{ $B_s^0-\overline{B}_s^0$ mixing and $B \to \pi K $ decays in stringy
leptophobic $Z^\prime$
}

\author{
Seungwon~Baek$^a$\footnote{sbaek@korea.ac.kr}, ~
Jong~Hun~Jeon$^b$\footnote{jben44@gmail.com}, ~ and
C.~S.~Kim$^b$\footnote{cskim@yonsei.ac.kr}
}

\affiliation{
$^a$
The Institute of Basic Science and Department of Physics,
Korea University, Seoul 136-701, Korea
\\
$^b$
Department of Physics, Yonsei University,
Seoul 120-479, Korea
}

\date{\today}

\begin{abstract}
\noindent
We consider a leptophobic $Z^\prime$ scenario in a flipped SU(5) grand unified theory
obtained from heterotic string theory. We show that the allowed $Z^\prime$ mass,
flavor conserving and flavor changing couplings of the $Z^\prime$ to the down-type quarks are
strongly constrained by the mass difference in $B_s-\overline{B}_s$ system and
the four branching ratios of $B \to \pi K$ decays.
It is shown that even under these constraints large deviations in
direct and/or indirect CP asymmetries of $B \to \pi K$ decays from the SM expectations are allowed.
Especially it is possible to accommodate the apparent puzzling data in $B \to \pi K$ CP asymmetries.
\end{abstract}

\maketitle
\section{Introduction}
The flavor changing neutral current (FCNC), in particular the
$\bar{b} \to \bar{s}$ transition, is a sensitive probe
of new physics (NP) beyond the standard model (SM) of particle physics.
The processes, such as $B \to X_s \gamma$~\cite{bsr}, $B \to \pi K$~\cite{B2piK},
$B \to \rho (\phi) K^*$~\cite{B2roK},
$B \to \phi K_S$~\cite{B2phiK}, and $B_s \to \mu^+ \mu^-$~\cite{Bs2mumu}
which are dominated by the $\bar{b} \to \bar{s}$ transiton have attracted much interest
because they still allow much room for large NP contributions.
The experimental data for some of them show apparent deviations from
the SM predictions~\cite{B2piK,B2roK,B2phiK}.

A viable NP scenario which may give large contribution to $\btos$ FCNC is
leptophobic $Z^\prime$~\cite{BJK_Zprime}. Extra U(1) gauge groups
appear naturally in many extensions of the SM.
If some of them remain unbroken down to the electroweak
scale, the $Z^\prime$ can be light and affect low energy phenomenology.
In addition, there is a possibility that the new neutral gauge boson does not couple to
leptons. This kind of  $Z^\prime$ gauge boson is called
{\it leptophobic}.
In case the $Z^\prime$ does not mix with the SM $Z$ boson,
the strong constraints from the electroweak precision
tests can be avoided.
Explicit leptophobic $Z^\prime$ model with these properties
has been constructed by Lopez, Nanopoulos and Yuan~\cite{LNY} in
heterotic string theory.

The model has gauge group,
$G = G_{\rm obs} \times G_{\rm hidden} \times G_{\rm U(1)}$,
where $G_{\rm obs} = {\rm SU(5)} \times {\rm U(1)}$,
$G_{\rm hidden} = {\rm SU(4)} \times {\rm SO(10)}$,
and $G_{\rm U(1)} = {\rm U(1)^5}$. It also has 63 massless matter
fields. It can be shown that the $Z^\prime$ gauge boson can be light
and leptophobic without mixing with $Z$.
The additional feature of this model is that the $Z^\prime$ coupling
is generation dependent. Therefore tree-level FCNC is generated
in general. Since $u^c$ and $L$ belong to the same multiplet
which do not couple to the leptophobic $Z^\prime$, the $Z^\prime$
coupling to $u$-quarks maximally violates parity.
And the disparity of the couplings to the $d^c$ and $u^c$
provides additional source of isospin breaking.

We study the contribution of leptophobic $Z^\prime$ to
$\BsBs$ mixing and non-leptonic decays $B \to \pi K$'s with this model
in mind. However, our analysis can be easily extended to other
(leptophobic) $Z^\prime$ models allowing tree-level FCNC's.

The measurement of
the mass difference, $\Delta m_s$, in the $B_s^0 -\overline{B}_s^0$ system
CDF~\cite{CDF_dms} collaborations
 \begin{eqnarray}
\Delta m_s^{\rm exp} = 17.77 \pm 0.10 (\text{stat}) \pm 0.07 (\text{syst})
\;\text{ps}^{-1}
 \end{eqnarray}
is consistent with the SM calculations~\cite{lattice_QCD}
\bea
\Delta m_s^{\rm SM}\Big|_{\rm (HP+JL)QCD} =
22.57_{-5.22}^{+5.88} ~ \text{ps}^{-1}.
\eea
This constrains many NP models~\cite{dms_th,dms_zprime,dms_MSSM,dms_etc}
including $Z^\prime$ models, MSSM models, {\it etc}.
In~\cite{BJK_Zprime}, we showed that the $\Delta m_s$ constraint
on the leptophobic $Z^\prime$ is much stronger than the
previously considered one~\cite{JKLY_Zprime} from the semi-leptonic
$B$-decays.
In this paper we extend the analysis in~\cite{BJK_Zprime} to
include the case where $Z^\prime$ couples to both left-handed
and right-handed quarks simultaneously. When both couplings
exist at the same time, we will see that the $\Delta m_s$ constraint
is not enough to set the upper bound on the sizes of the FCNC couplings.
We go beyond the \cite{BJK_Zprime} and demonstrate
that the additional constraints are available,
{\it i.e.} the four branching ratios (BR) of $B \to \pi K$'s, which
can give the limits even in the simultaneous existence of both left-
and right-handed couplings.

The charmless non-leptonic decays $B \to \pi K$ have been measured
precisely enough to probe the electroweak amplitudes~\cite{B2piK}.
The experimental data indicate that while the BRs are consistent with the SM
expectations, some of direct and indirect CP asymmetries
 show apparent (but still debatable) deviations from the
SM~\cite{BL_B2piK}.
Accepting this discrepancy seriously,
we can see the electroweak penguin
sector is the best place to search for NP~\cite{BL_B2piK}.
We will see that in our model the
leptophobic $Z^\prime$ can give large contributions to the electroweak
penguin amplitudes while satisfying the $\Delta m_s$ and the
$BR(B \to \pi K)$'s. In addition we will see that the predicted
direct and indirect CP asymmetries can accommodate the
discrepancies between the SM predictions and measurements simultaneously.

We note that the merit of our model in explaining the data comes from
(i) it automatically evades the stringent constraints
involving leptons, such as LEP I data, $B_s \to \mu^+ \mu^-$,
(ii) there are new CP violating phases, and
(iii) the characteristic isospin breaking interaction in this model
 can generate large electroweak penguins.

The paper is organized as follows:
In section~\ref{sec:model} we briefly describe our model, the leptophobic
$Z^\prime$ model in the stringy flipped SU(5) theory.
In section~\ref{sec:BsBs} we calculate the $Z^\prime$
contribution to the $\Delta m_s$ when $Z'$ couplings to the quarks
have both handedness.
In section~\ref{sec:B2piK} we consider the constraints imposed
by the $BR(B \to \pi K)$ and predict the deviations in the direct
and indirect CP asymmetries from the SM expectation and
compare with the experimental results.
We conclude in section~\ref{sec:conclusion}.

\section{The Model}
\label{sec:model}

The leptophobic $Z^\prime$ model
can occur naturally in some grand unified theories (GUTs) and
string theories. It naturally avoids
stringent low energy constraints thanks to the absence of couplings to
charged leptons and light neutrinos.
There are at least two known mechanisms that can generate the
leptophobic $Z^\prime$.
The first one is obtained via dynamical mixing between the U(1)
and U(1)$'$ in the $E_6$ GUT~\cite{E6}.
The other scenario of leptophobia is obtained in the stringy
flipped SU(5) GUT in the heterotic string theory~\cite{LNY}.
In this paper we consider only the latter scenario only because it
is more relevant to the $B \to \pi K$ decays.

In the flipped SU(5), the SM particles appear in three copies
of the representations
\bea
 F=( {\bf 10}, {1 \over 2}) = \{Q, d^c, \nu^c \},
 \quad \ol{f} = ( \ol{\bf 5}, -{3 \over 2} ) =\{L, u^c \},
 \quad \ell^c = ( {\bf 1},{5 \over 2}) = \{e^c \}.
\eea
The new neutral gauge boson $Z^\prime$ can be leptophobic if
it does not couple to $\ol{\bf 5}$ and ${\bf 1}$, while the quarks
in ${\bf 10}$ still couple to it~\cite{LNY}.

In addition to its own beauty this scenario has the following
phenomenologically
interesting features:
\begin{itemize}
\item The new $Z^\prime$ coupling is generation dependent and can
generate FCNC processes.
\item The FCNC couplings allow large CP violation.
\item It violates the isospin symmetry in the right-handed up- and down-quarks.
\item The new gauge boson interaction maximally violates the parity
in the up-quark
sector.
\end{itemize}
In the mass eigenstates the interactions of $Z^\prime$ gauge boson
with the quarks
can be written as
\bea
 {\cal L} = - {g_2 \over \cos\theta_W} \de Z^\prime_\mu \Bigg(
 \ol{u} \ga^\mu P_L \Big[V_L^u \hat{c} V_L^{u\dagger} \Big] u
+\ol{d} \ga^\mu P_L \Big[V_L^d \hat{c} V_L^{d\dagger} \Big] d
+\ol{d} \ga^\mu P_R \Big[V_R^d \hat{c} V_R^{d\dagger} \Big] d \Bigg),
\label{eq:Zp-q-q}
\eea
where $\de$ parameterizes the size of the new gauge coupling
relative to the SM coupling and is expected to be of ${\cal O}(1)$.
The $\hat{c} = {\rm diag} (c_1, c_2, c_3)$ represent the generation-dependent
U(1)$'$ quantum numbers~\cite{LNY}, and
$V_{L,R}^q (q=u,d)$ are unitary matrices diagonalizing the quark mass matrices.
The explicit sets of values for $c_i (i=1,2,3)$ derived from heterotic
string theory are given in~\cite{LNY}.
Since $V_L^u, V_L^d, V_R^d$ are unknown, we do not take specific
values of $c_i$'s in~\cite{LNY} and take them as free parameters.

We introduce complex parameters, $L$ and $R$,
\bea
 \Big[V_L^d \hat{c} V_L^{d\dagger} \Big]_{23} \equiv \frac{1}{2} L_{sb}^{Z'}, \quad
 \Big[V_R^d \hat{c} V_R^{d\dagger} \Big]_{23} \equiv \frac{1}{2} R_{sb}^{Z'}.
 \label{eq:LR}
\eea
to represent the
$b \to s$ FCNC couplings.
For comparison with~\cite{BJK_Zprime}, we kept the factor 2 in
(\ref{eq:LR}).
Then
\bea
 {\cal L}^{Z'}_{\rm FCNC}
 &=& -{g_2 \over 2 \cos\theta_W}
\l[ L_{sb}^{Z'} \ol{s}_L \ga_\mu b_L Z^{\prime \mu}
  + R_{sb}^{Z'} \ol{s}_R \ga_\mu b_R Z^{\prime \mu}\r] + h.c,
\eea
where $\delta$ in (\ref{eq:Zp-q-q}) is absorbed into $L$ and $R$.
To calculate the $B \to \pi K$ decay amplitudes we also need
the $Z^\prime$ couplings to the first generation quarks.
Although $Q$ and  $d^c$ have the same $U^\prime(1)$
charges, the mixing effect in (\ref{eq:Zp-q-q}) can give
different couplings to the $Z^\prime$ in general
\bea
 {\cal L}(Z^\prime \ol{q} q) &=& -{g_2 \over  \cos \theta_W} \de  \;   Z^{'\mu}
 \l[ \ol{u} \ga_\mu c_L^u P_L u
     +  \ol{d} \ga_\mu (c_L^d P_L + c_R^d P_R) d
 \r],
\label{eq:Zprime_qq}
\eea
where we defined
\bea
 c_L^u \equiv \Big[V_L^u \hat{c} V_L^{u\dagger} \Big]_{11}, \quad
 c_L^d \equiv \Big[V_L^d \hat{c} V_L^{d\dagger} \Big]_{11}, \quad
 c_R^d \equiv \Big[V_R^d \hat{c} V_R^{d\dagger} \Big]_{11}.
\eea
From the structure of CKM matrix, we assume the couplings to
the left-handed quarks are approximately equal, {\it i.e.}
$c_L^u = c_L^d  \equiv c_L^q$.
However, in general $c_R^d$ can be different from $c_L^q$.
As mentioned above, the absence of $c_R^u$ is
the characteristic feature of the leptophobic flipped SU(5) scenario.
Note that since $\de$ and $c_L^q$ are unknown, $c_L^q$ can always be absorbed
to $\de$. In addition, $\de$ does not appear in the expression for
$\Delta m_s$
and it can absorbed into
$L_{sb}^{Z'}$ or $R_{sb}^{Z'}$ in the $B \to \pi K$ amplitudes.
So we fix $\de = c_L^q = 1$ from now on.

\section{$\BsBs$ Mixing}
\label{sec:BsBs}

In general the $Z'$ can couple to both left- and
right-handed quarks simultaneously as can be seen in (\ref{eq:Zp-q-q}).
Then we need to extend the operator basis beyond the SM one in the effective
Hamiltonian describing $\BsBs$ mixing.

The most general $\Delta B =\Delta S=2$ process
is described by the effective Hamiltonian~\cite{Becirevic:2001jj}:
\begin{eqnarray}
{\cal{H}}_{eff} = \sum_{i=1}^5 C_i Q_i
+\sum_{i=1}^3\tilde{C}_i \tilde{Q}_i
+ h.c,
\end{eqnarray}
where
\begin{eqnarray}
Q_1&=&\bar{s}_L^\alpha \gamma_\mu b_L^\alpha
      \bar{s}_L^\beta \gamma^\mu b_L^\beta
\nonumber\\
Q_2&=&\bar{s}_R^\alpha b_L^\alpha
      \bar{s}_R^\beta b_L^\beta
\nonumber\\
Q_3&=&\bar{s}_R^\alpha b_L^\beta
      \bar{s}_R^\beta b_L^\alpha
\nonumber\\
Q_4&=&\bar{s}_R^\alpha b_L^\alpha
      \bar{q}_L^\beta b_R^\beta
\nonumber\\
Q_5&=&\bar{s}_R^\alpha b_L^\beta
      \bar{s}_L^\beta b_R^\alpha
\end{eqnarray}
and the operators $\widetilde{Q}_{1,2,3}$ are obtained from the $Q_{1,2,3}$ by
the exchange $L \leftrightarrow R$. Here $q_{R,L}=P_{R,L}\,q$, with
$P_{R,L}=(1 \pm \gamma_5)/2$, and $\alpha$ and $\beta$ are color indices.

In our model the nonvanishing Wilson coefficients at $M_{Z^\prime}$ scale are simply
given by
\bea
 C_1(M_{Z^\prime}) &=& \frac{g_Z^2}{8 M_{Z^\prime}^2}
\l({L_{sb}^{Z^\prime}}\r)^2,
\nonumber\\
\widetilde{C}_1(M_{Z^\prime}) &=& \frac{g_Z^2}{8 M_{Z^\prime}^2}
\l({R_{sb}^{Z^\prime}}\r)^2,
\nonumber\\
C_5(M_{Z^\prime}) &=& \frac{g_Z^2}{8 M_{Z^\prime}^2}
         \l(-2 {L_{sb}^{Z^\prime}} {R_{sb}^{Z^\prime}}\r).
\eea
Here $Q_5$ is the additionally generated operator compared with~\cite{BJK_Zprime}.
The renormalization group running down to $m_b$ scale mixes $Q_5$ with
$Q_4$ and we get~\cite{Becirevic:2001jj}
\bea
 C_1(\mu_b) & \simeq& 0.801~C_1(M_{Z^\prime}),
\nonumber\\
C_4(\mu_b) & \simeq& 0.697~C_5(M_{Z^\prime}),
\nonumber\\
C_5(\mu_b) & \simeq& 0.886~C_5(M_{Z^\prime}).
\eea
The other operators are not generated at all and
we get $C_2(\mu_b) = C_3(\mu_b) = 0$.

Now we can calculate the $\BsBs$ mixing matrix element
\bea
 M^s_{12} = M^{s, {\rm SM}}_{12} + M^{s, Z^\prime}_{12}
        \equiv M^{s, {\rm SM}}_{12} (1 + R),
\eea
where $R \equiv M^{s, Z^\prime}_{12} / M^{s, {\rm SM}}_{12}$.
The SM contribution $M^{s, {\rm SM}}_{12}$ is given by~\cite{Buras:1990fn},
\bea
 M_{12}^{s,\rm SM} = {G_F^2 M_W^2 \over 12 \pi^2}
  M_{B_s} \l(f_{B_s} \hat{B}_{B_s}^{1/2} \r)^2
  \eta_B S_0(x_t) \l(V_{tb} V_{ts}^*\r)^2,
\label{eq:M12_SM}
\eea
where
$\hat{B}_{B_s} \simeq
B_1(\mu_b) [\al_s(\mu_b)]^{-6/23} \l[1+1.627 \al_s(\mu_b) /( 4 \pi) \r]$.
The $Z^\prime$ contribution
\bea
 M^{s, Z^\prime}_{12} &=& {1 \over 3} M_{B_s} f_{B_s}^2 \l[
   \Big(C_1(\mu_b) + \wt{C}_1(\mu_b) \Big)  B_1(\mu_b) \right. \nl
  && \l. +{1 \over 4} \;\l( M_{B_s} \over m_b(\mu_b) + m_s(\mu_b) \r)^2
   \Big(  3 C_4(\mu_b) B_4(\mu_b) +  C_5(\mu_b) B_5(\mu_b)\Big)
  \r]
\eea
involves additional hadronic parameters, $B_4(\mu_b)$ and $B_5(\mu_b)$.

The mass difference in the $\BsBs$ system, $\Delta m_s$ is obtained by
\bea
   \Delta m_s = 2 |M_{12}^s|.
\eea
In the SM, we get
\bea
  \Delta m_s^{\rm SM} = (22.5 \pm 5.5)\;\; \mbox{ps}^{-1},
\eea
where the nonperturbative hadronic parameters $f_{B_s}$ and  $\hat{B}_{B_s}$ are
the main sources of the uncertainty. We used the value
\bea
  f_{B_s} \hat{B}_{B_s}^{1/2} \Bigg|_{\rm (HP+JL)QCD} = (0.295 \pm 0.036) \;\; {\rm GeV},
\eea
which is the combined lattice result~\cite{lattice_QCD}
from JLQCD and HPQCD.
For other parameters, we used $\al_s(\mu_b)=0.22$, $\eta_B = 0.551$,
$\ol{m}^{\ol{MS}}_t(m_t) = 162.3$ GeV and
$V_{ts}=0.04113$~\cite{Charles:2004jd}.

\begin{figure}
\begin{center}
\includegraphics[width=0.3\textwidth]{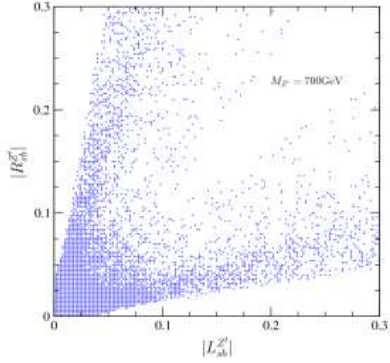}
\end{center}
\caption{
The allowed region in ($\l|L_{sb}^{Z'}\r|$, $\l|R_{sb}^{Z'}\r|$) plane by $\De m_s$
alone.
}
\label{fig:LR}
\end{figure}
In Figure~\ref{fig:LR} the allowed region by $\Delta m_s$ alone is shown. We fixed
$m_{Z^\prime} = 700 {\rm GeV}$ which
is above the experimental lower bound~\cite{Zprime_mass} and scanned the weak phases
$\phi_{L(R)}^{Z^\prime} \equiv \arg(L_{sb}^{Z'} (R_{sb}^{Z'}))$
from 0 to $2 \pi$ independently.
We can see that the sizes of FCNC couplings, $\l|L_{sb}^{Z'}\r|$
and $\l|R_{sb}^{Z'}\r|$,
are restricted typically to be less than $\sim$ 0.1 for most values of the
scanned parameters, which is consistent with~\cite{BJK_Zprime}.
However, the two additional bands appear in this case due to the
cancelation between $L_{sb}^{Z^\prime}$ and $R_{sb}^{Z^\prime}$.
These regions extend indefinitely and show that the $\Delta m_s$ alone
is not enough to constrain both $L_{sb}^{Z^\prime}$ and $R_{sb}^{Z^\prime}$
simultaneously. We will show that the
$BR(B \to \pi K)$ can give upper bounds on both
$|L_{sb}^{Z^\prime}|$'s and $|R_{sb}^{Z^\prime}|$ in the next section.

\section{$B \to \pi K$ Decays}
\label{sec:B2piK}

The $B \to \pi K$ decays are dominated by the $\btos$ QCD penguin
diagrams. The subdominant electroweak penguin contribution is also
sizable and may play important role in probing the NP as mentioned
in the Introduction.
The current experimental data in Table~\ref{tab:b2pik}
show the branching ratios are quite precisely measured and
the so-called $R_c/R_n$ puzzle~\cite{Buras} has disappeared.
Therefore we take the four BRs as the additional constraints
to the $\Delta m_s$ constraint considered in section~\ref{sec:BsBs}.
As we will see in a moment, they are orthogonal
to and as strong as  $\Delta m_s$ constraint.

\begin{table}[tbh]
\center
\begin{tabular}{cccc}
\hline
\hline
Mode & $BR[10^{-6}]$ & $A_{\rm CP}$ & $S_{\rm CP}$ \\ \hline
$B^+ \to \pi^+ K^0$ & $23.1 \pm 1.0$ & $0.009 \pm 0.025$ & \\
$B^+ \to \pi^0 K^+$ & $12.9 \pm 0.6$ & $0.050 \pm 0.025$ & \\
$B^0 \to \pi^- K^+$ & $19.4 \pm 0.6$ & $-0.097 \pm 0.012$ & \\
$B^0 \to \pi^0 K^0$ & $9.9 \pm 0.6$ & $-0.14 \pm 0.11$ &
$0.38 \pm 0.19$ \\
\hline
\hline
\end{tabular}
\caption{Branching ratios, direct CP asymmetries $A_{\rm CP}$, and
mixing-induced CP asymmetry $S_{\rm CP}$ (if applicable) for the four
$\btopik$ decay modes. The data are taken from Refs.~\cite{HFAG} and
\cite{piKrefs}.}
\label{tab:b2pik}
\end{table}

In the SM the $B \to \pi K$ decay amplitudes can be written
in terms of topological
amplitudes:
\bea
A(B^+ \to \pi^+ K^0) &=& -P^{'}_{tc} -{1 \over 3} P^{'C}_{\rm EW}
+ P^{'}_{uc} e^{i\gamma} , \nl
\sqrt{2} A(B^+ \to \pi^0 K^+) &=& P^{'}_{tc} - P'_{\rm EW}
-{2 \over 3} P^{'C}_{\rm EW} - \l(T' + C' + P^{'}_{uc} \r) e^{i\gamma} , \nl
A({B}^0 \to \pi^- K^+) &=& P^{'}_{tc}  -{2 \over 3} P^{'C}_{\rm EW}
   -\l(T' + P^{'}_{uc}\r) e^{i\gamma}, \nl
\sqrt{2} A({B}^0 \to \pi^0 {K}^0) &=&
-P^{'}_{tc} - P'_{\rm EW}  -{1 \over 3} P^{'C}_{\rm EW}
-\l(C' - P^{'}_{uc}\r) e^{i\gamma},
\label{eq:all2}
\eea
where other small annihilation and exchange amplitudes are neglected.
Here the weak phase, $\gamma$, dependence has been explicitly written.
The primes denote the $\btos$ transition.
The $P^{'}_{tc}$ ($P^{'}_{uc}$) is the QCD penguin amplitude with
$t,c$ ($u, c$) quarks running inside the loop. The tree (color-suppressed tree)
diagrams are represented by $T'$ ($C'$). The $P^{'(C)}_{\rm EW}$ is the
electroweak (color-suppressed electroweak) penguins and related to
the $T'(C')$ by flavor SU(3) symmetry~\cite{GPY}:
\bea
P'_{\rm EW} &=& {3 \over 4} {C_9 + C_{10} \over C_1 + C_2} R(T' + C')
+ {3 \over 4} {C_9 - C_{10} \over C_1 - C_2} R(T' - C'), \nl
P^{'C}_{\rm EW} &=& {3 \over 4} {C_9 + C_{10} \over C_1 + c_2} R(T' + C')
- {3 \over 4} {C_9 - c_{10} \over C_1 - C_2} R(T' - C'),
\label{eq:GPY}
\eea
where $C_i$ (i=1,2,9,10) are the Wilson coefficients
and $R = \l|V_{ts} V^*_{tb} / V_{us} V^*_{ub}\r| $.

In the SM, from the loop-, color-factor and the hierarchy of CKM matrix elements,
we expect the following hierarchies:
\bea
\begin{array}{cc}
O(1)   & |P'_{tc}|, \\
O(\bar{\lambda})  & |T'|, |P'_{\rm EW}|, \\
O(\bar{\lambda}^2)  & |C'|, |P'_{uc}|,|P^{'C}_{\rm EW}|, \\
O(\bar{\lambda}^3)  & |A'|.
\label{eq:hierarchy}
\end{array}
\eea

However the experimental data in Table~\ref{tab:b2pik} are not fully
consistent with these hierarchies. Specifically $A_{\rm CP}(B^+ \to
\pi^0 K^+) \not \simeq A_{\rm CP}(B^0 \to \pi^- K^+)$ and $S_{\rm
CP}(B^0 \to \pi^0 K^0) \not \simeq \sin 2 \beta$ require
$|C'/T'|=1.6 \pm 0.3$~\cite{BL_B2piK}. This large ratio is
inconsistent with the SM expectation (\ref{eq:hierarchy}) which is
supported by theoretical calculations~\cite{QCDF,PQCD_NLO,SCET}. And
the implications of this apparent discrepancy have been considered
in many NP models~\cite{B2piK}. This puzzle can be solved most
naturally if NP is introduced in the electroweak penguin
amplitude~\cite{BL_B2piK}.

In this paper, as mentioned above, we use the four $BR(B \to \pi K)$'s to constrain the
sizes of $Z'$ couplings.
Using the remaining parameters after imposing $\Delta m_s$
and the $BR(B \to \pi K)$'s, we predict the direct and indirect
CP asymmetries which show apparent deviations from the SM.
We show that the predictions for
$A_{\rm CP}(B^+ \to \pi^0 K^+)$,
$A_{\rm CP}(B^0 \to \pi^- K^+)$ and
$S_{\rm CP}(B^0 \to \pi^0 K^0)$
in our model are in the right directions to the experimental
results.

To calculate the SM predictions for the BR's and CP asymmetries, we
use the NLO calculations in the perturbative QCD (PQCD) results~\cite{PQCD_NLO}.
In our model the $Z^\prime$ contributions can change the amplitudes,
$P'_{tc(uc)}$, $P^{'(C)}_{\rm EW}$.
The NP contributions are calculated using the naive factorization
method and their strong phases are assumed to be equal to
the corresponding SM diagrams.
We will also discuss the effect of the NP strong phase later.
The $Z^\prime$ contribution to
the topological amplitudes are written in terms of the Wilson coefficients
in the standard operator basis~\cite{BBL} as
\bea
P'(Z') &=& -\la_t \Big[(a_4 + r^K_\chi a_6)
-(\tilde{a}_4 + r^K_\chi \tilde{a}_6) \Big]
A_{\pi K} e^{i \de'_{tc}}\nl
P'_{\rm EW}(Z') &=&
\frac{3}{2} \la_t \Big[(-a_7 +  a_9)
-(-\tilde{a}_7 + \tilde{a}_9) \Big] A_{K \pi} e^{i \de'_{\rm EW}}\nl
P^{'C}_{\rm EW}(Z') &=& \frac{3}{2} \la_t
\Big[(a_{10} + r^K_\chi a_8)
-(\tilde{a}_{10} + r^K_\chi \tilde{a}_8)
\Big]
A_{\pi K} e^{i \de^{'C}_{\rm EW}},
\eea
where $\la_t = V_{ts}^* V_{tb}$, $a_i = C_i + C_{i\pm 1}/3$ ($+(-)$ for
odd (even) $i$), $r^K_\chi = 2 m_K^2/m_b(m_s + m_q)$
(with $m_q = (m_u + m_d)/2$),
$A_{\pi K (K \pi)} = G_F (m_B^2 - m_\pi^2) F_0^{\pi (K)} f_{K (\pi)}/\sqrt{2}$,
and the $\de$'s are the corresponding strong phases obtained
in \cite{PQCD_NLO}. The $\tilde{a}_i$'s are Wilson coefficients
for the chirality flipped operators.
In our model the Wilson coefficients at $M_{Z^\prime}$ scale are
\bea
  -\la_t C_3(M_{Z'}) = \de \frac{m_Z^2}{m_{Z'}^2} L_{sb}^{Z'}
  \frac{c_L^u + 2 c_L^d}{3}, &\quad&
  C_4(M_{Z'}) = 0, \nl
  -\la_t C_5(M_{Z'}) = \de \frac{m_Z^2}{m_{Z'}^2} L_{sb}^{Z'}
  \frac{c_R^u + 2 c_R^d}{3}, &\quad&
  C_6(M_{Z'}) = 0, \nl
  -\frac{3}{2} \la_t C_7(M_{Z'}) = \de \frac{m_Z^2}{m_{Z'}^2} L_{sb}^{Z'}
  (c_R^u - c_R^d), &\quad&
 C_8(M_{Z'}) = 0, \nl
  -\frac{3}{2} \la_t C_9(M_{Z'}) = \de \frac{m_Z^2}{m_{Z'}^2} L_{sb}^{Z'}
  (c_L^u - c_L^d), &\quad&
 C_{10}(M_{Z'}) = 0.
\label{eq:Zprime-WC}
\eea
There are also the chirality flipped operators to the SM operators.
Their Wilson coefficients, $\tilde{C}_i$'s are obtained by exchanging
$L_{sb}^{Z^\prime} \leftrightarrow R_{sb}^{Z^\prime}$,
$c_L^q \leftrightarrow c_R^q$.

\begin{figure}[bt]
\begin{center}
\subfigure[]{
\includegraphics[width=0.3\textwidth]{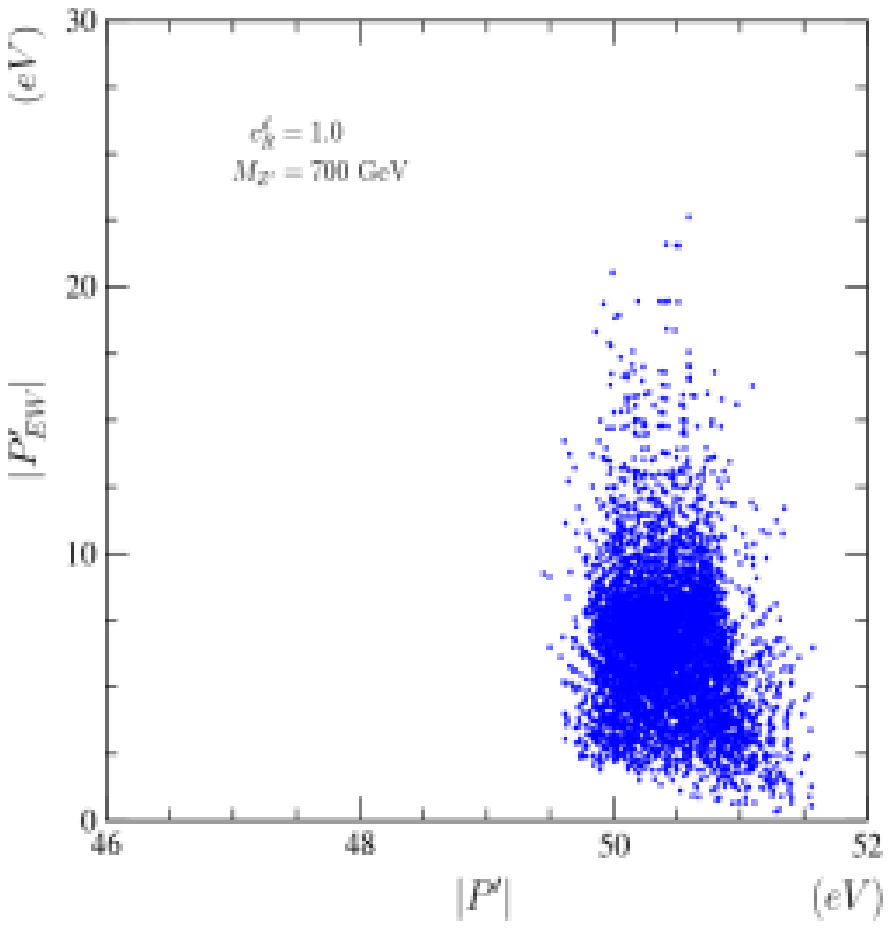}
}
\hspace{0.5cm}
\subfigure[]{
\includegraphics[width=0.3\textwidth]{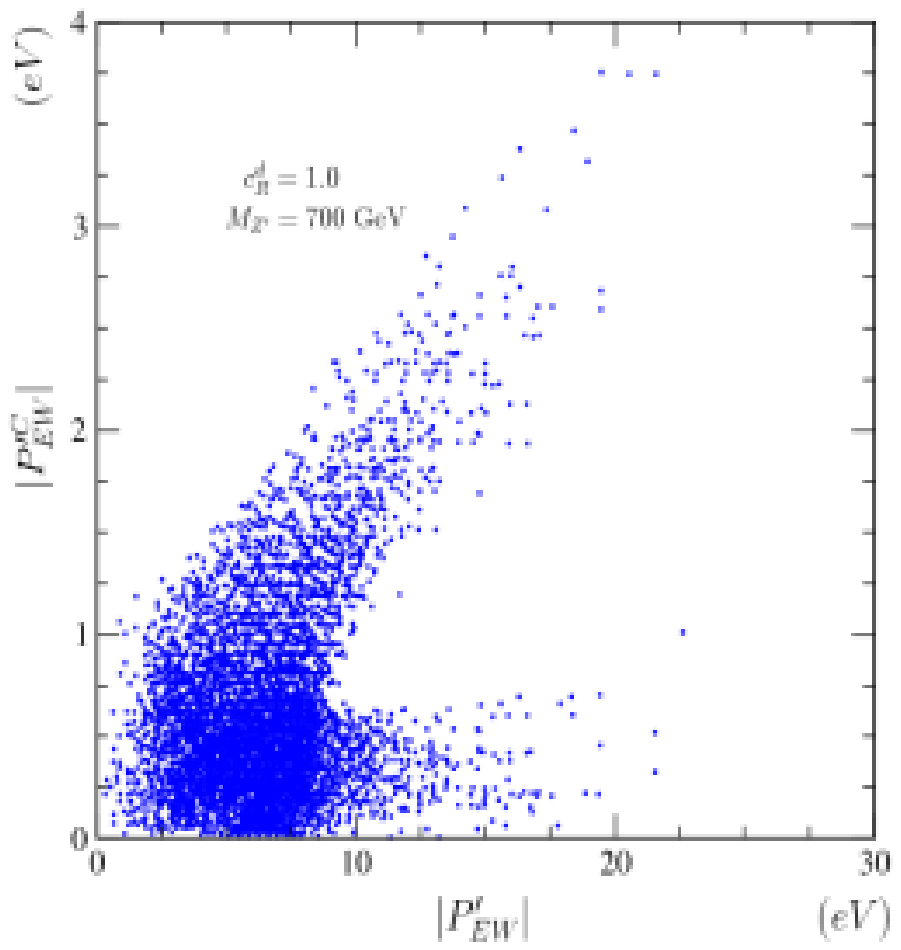}
}
\end{center}
\caption{
The correlations between
$P'_{tc}$ and $P'_{\rm EW}$ (a)
and between
$P'_{EW}$ and $P^{'C}_{\rm EW}$ (b) for $M_{Z'}=700$ GeV and $c_R^d=1$.
}
\label{fig:P_PEW}
\vspace{8pt}
\end{figure}
In Figure~\ref{fig:P_PEW}, we show the allowed topological
amplitudes
in ($P'_{tc}$, $P'_{\rm EW}$) plane
(a) and in ($P'_{EW}$, $P^{'C}_{\rm EW}$) plane (b).
We can see that $P'_{tc}$
is strongly constrained to lie in the region (49,52) eV by the
$BR(B \to \pi K)$'s,
whereas sizable deviation from the SM predictions are possible for
$P'_{\rm EW}$ and $P^{'C}_{\rm EW}$.

\begin{figure}[bt]
\begin{center}
\subfigure[]{
\includegraphics[width=0.26\textwidth]{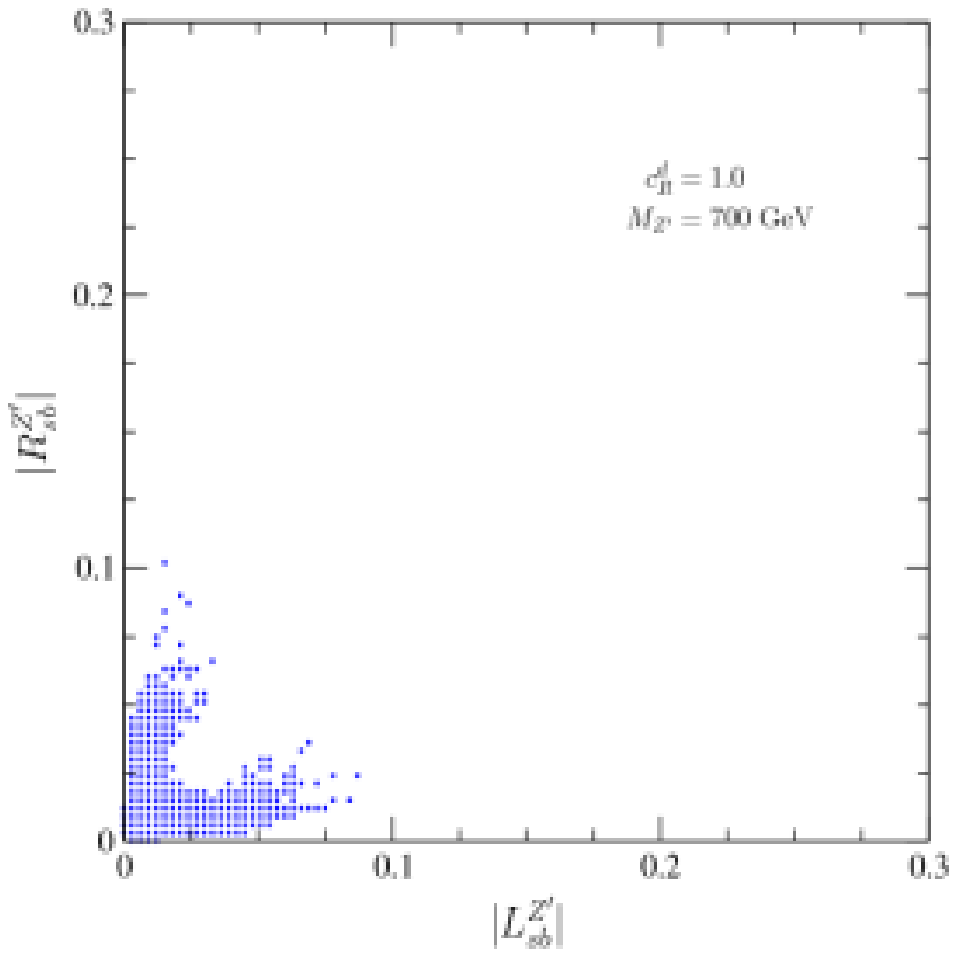}
}
\subfigure[]{
\includegraphics[width=0.25\textwidth]{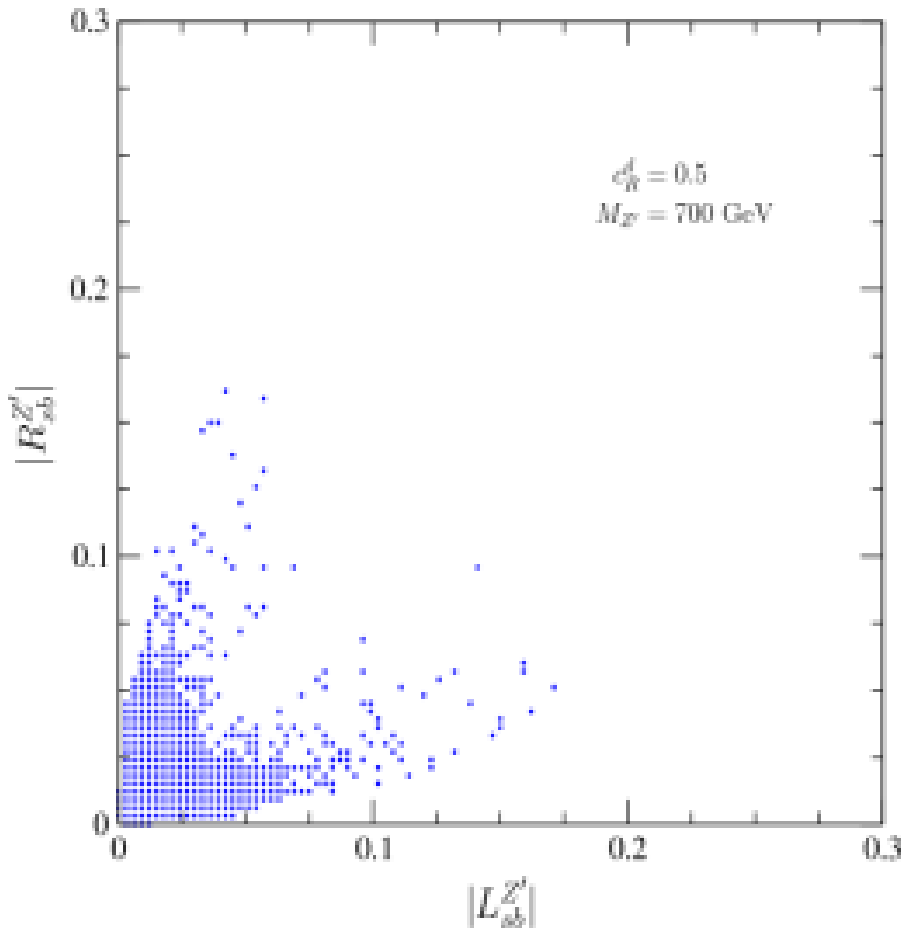}
}
\subfigure[]{
\includegraphics[width=0.25\textwidth]{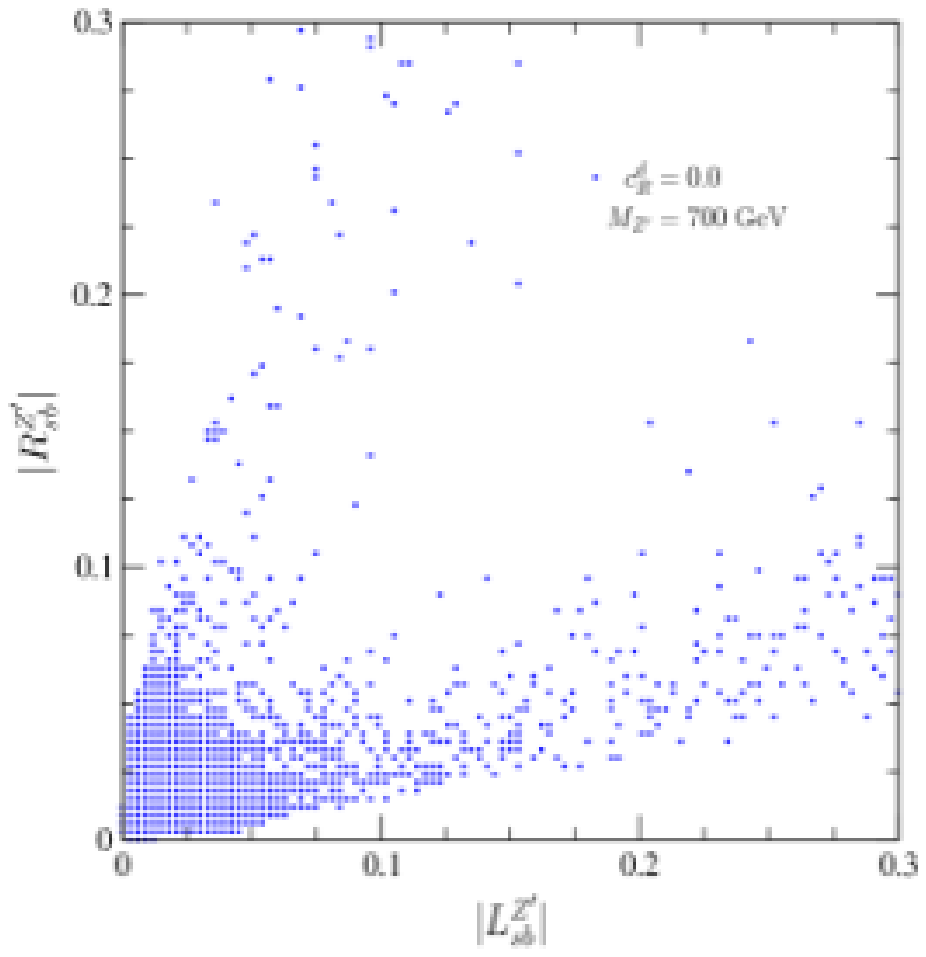}
}
\end{center}
\caption{
The allowed region in ($\l|L_{sb}^{Z'}\r|$, $\l|R_{sb}^{Z'}\r|$)
plane by $\De m_s$
and the four $BR(B \to \pi K)$'s.
We fixed $c_R^d=1.0, 0.5, 0.0$ from the left.
}
\label{fig:LR_B2piK}
\end{figure}
As mentioned in the previous section, the Figure~\ref{fig:LR_B2piK}
shows that, given $c_L^q$ and $c_R^d$,
 the flavor changing couplings, $|L_{sb}^{Z^\prime}|$
and $|R_{sb}^{Z^\prime}|$, are constrained by the four $BR(B \to \pi
K)$'s in addition to the $\Delta m_s$. Since the experimental
measurements are quite precise now and the theoretical calculations
have still large errors, we allowed 3-$\sigma$ range for the BRs.
For the plot, we set $M_{Z'} = 700$ GeV and $c_R^d =1, 0.5, 0$ from
the left, respectively. Since the $BR(B \to \pi K)$ decays are most
sensitive to the $P'_{\rm EW}$ which is maximized at $c_R^d=1$ and
vanishes at $c_R^d=0$, the constraint is strongest for $c_R^d=1$.

Now we predict the direct and indirect CP asymmetries using the parameter set
allowed by $\Delta m_s$ and the four $BR(B \to \pi K)$'s. We are especially
interested in the correlation between the two direct CP asymmetries,
$A_{\rm CP}(B^+ \to \pi^0 K^+)$ and $A_{\rm CP}(B^0 \to \pi^- K^+)$, and
the indirect CP asymmetry, $S_{\rm CP}(B^0 \to \pi^0 K^0)$ because they
show the apparent deviations from the SM predictions.

\begin{figure}[bt]
\begin{center}
\subfigure[]{
\includegraphics[width=0.25\textwidth]{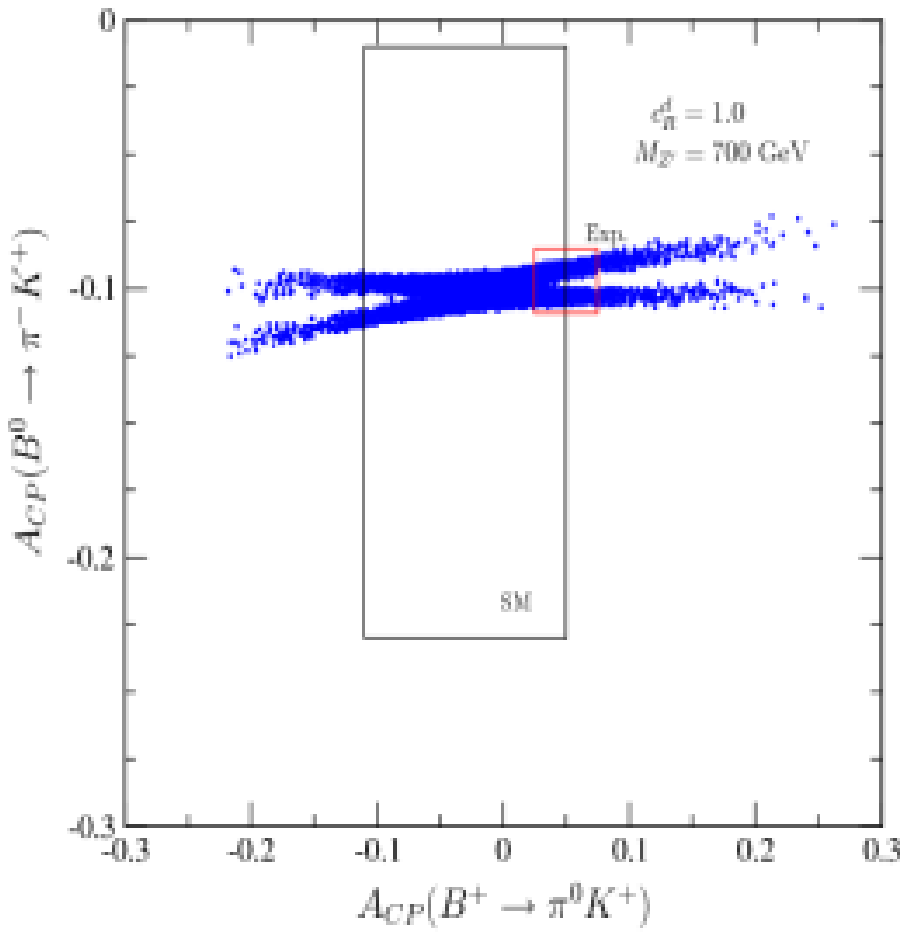}
}
\hspace{0.5cm}
\subfigure[]{
\includegraphics[width=0.25\textwidth]{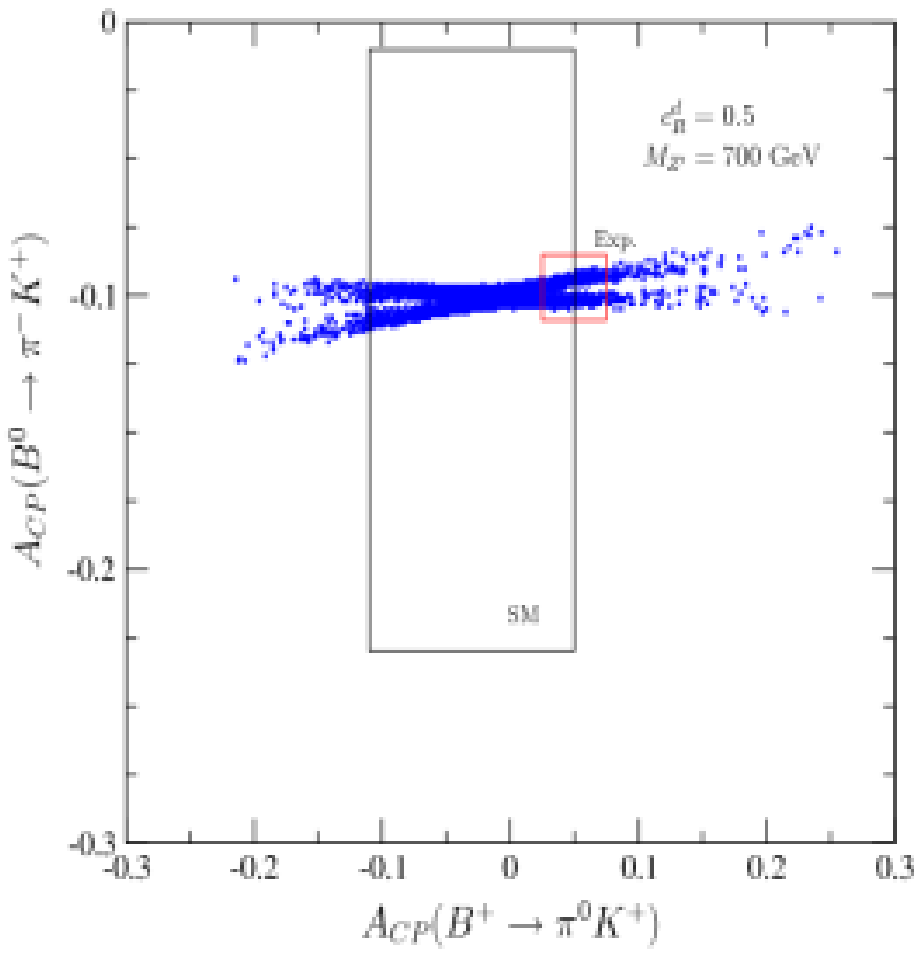}
}
\hspace{0.5cm}
\subfigure[]{
\includegraphics[width=0.26\textwidth]{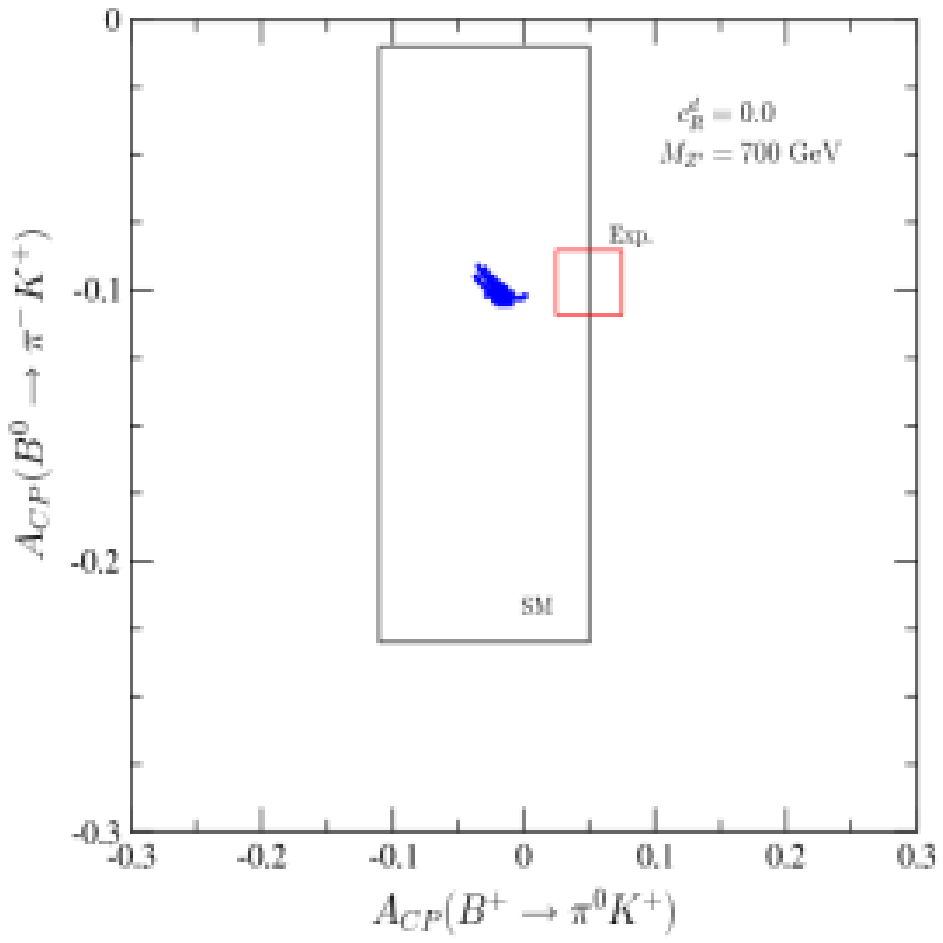}
}
\end{center}
\caption{
The predictions for
$A_{\rm CP}(B^+ \to \pi^0 K^+)$ and $A_{\rm CP}(B^0 \to \pi^- K^+)$ for $M_{Z'}=700$ GeV
and (a) $c_R^d=1.0$,
(b) $c_R^d=0.5$,
(c) $c_R^d=0.0$.
}
\label{fig:ACP}
\vspace{8pt}
\end{figure}

The predictions for the $A_{\rm CP}(B^+ \to \pi^0 K^+)$ and
$A_{\rm CP}(B^0 \to \pi^- K^+)$
are shown in Figure~\ref{fig:ACP} for $M_{Z'}=700$ GeV.
In these figures the errors for the SM predictions
are also obtained from \cite{PQCD_NLO}.
For the NP predictions we fixed the SM to
the central values and we did not include the hadronic uncertainties.
Although the SM results are consistent with the
experimental data at 2-$\sigma$ level,
the $Z^\prime$ contribution can accommodate
the current data at 1-$\sigma$ level for $c_R^d=1.0, 0.5$.
The value $c_R^d=0.0$ cannot
explain the data. These results are consistent
with~\cite{BL_B2piK} which claims
that the current data require large NP contributions
at the electroweak penguin
sector.

\begin{figure}
\begin{center}
\subfigure[]{
\includegraphics[width=0.25\textwidth]{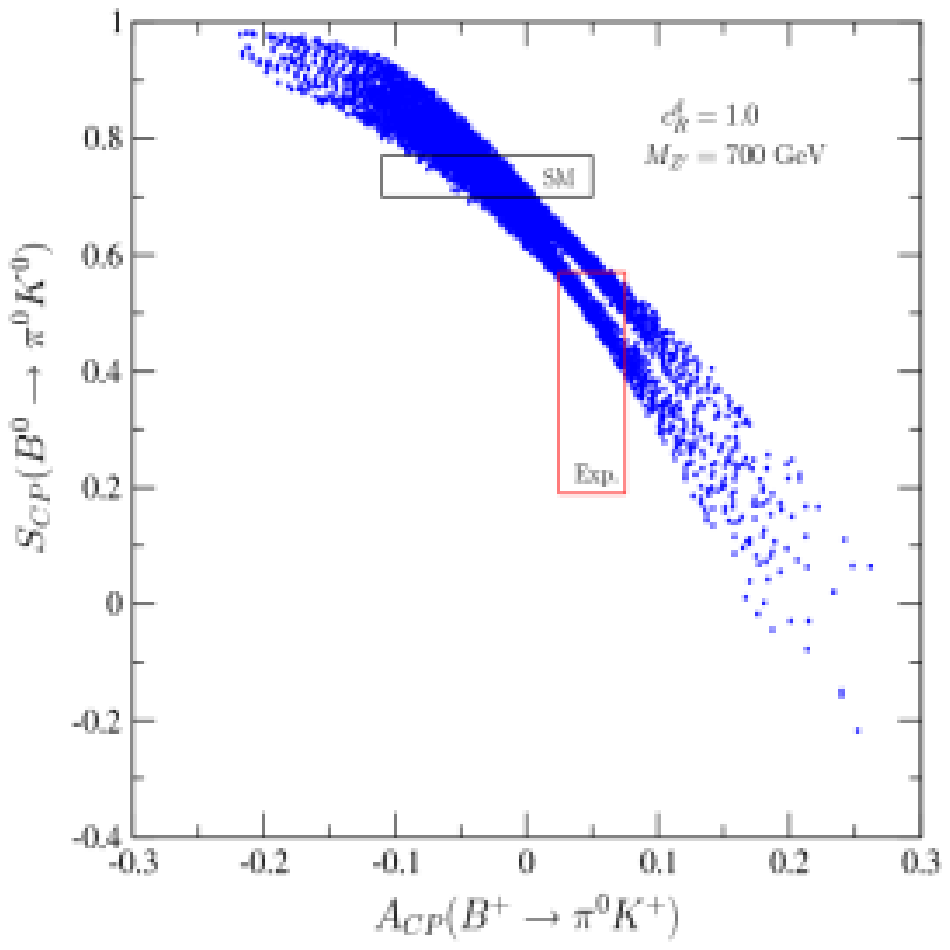}
}
\hspace{0.5cm}
\subfigure[]{
\includegraphics[width=0.25\textwidth]{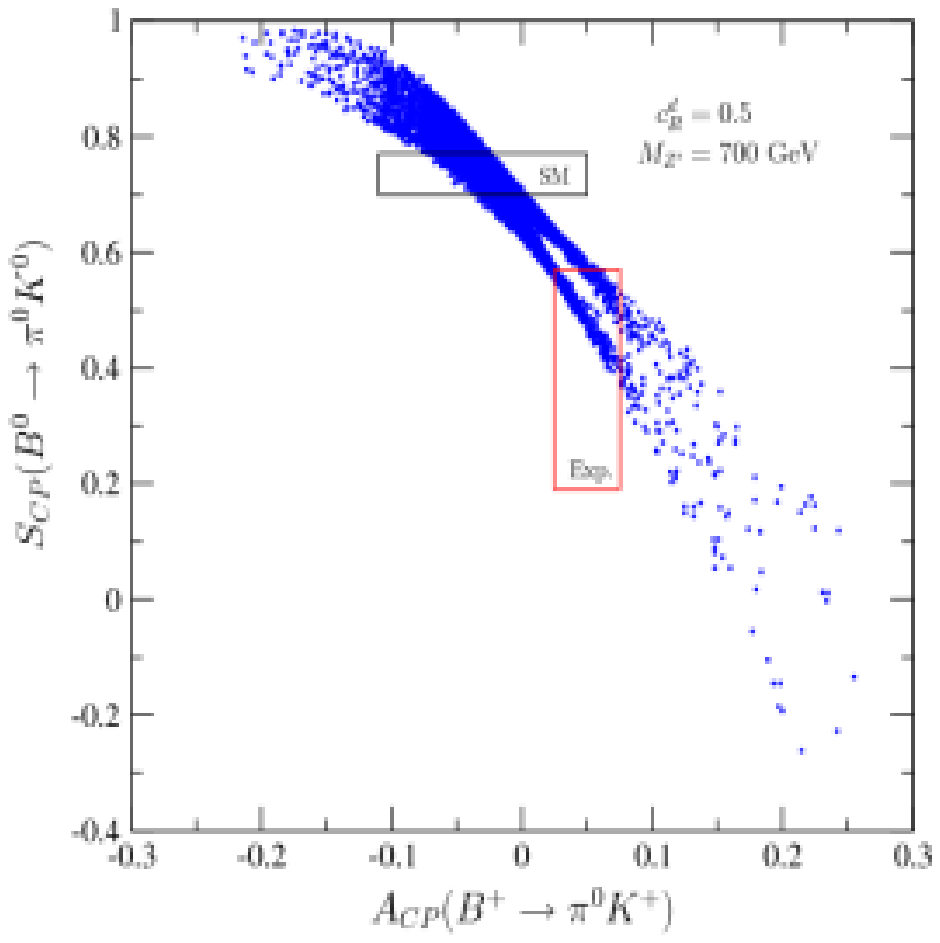}
}
\hspace{0.5cm}
\subfigure[]{
\includegraphics[width=0.25\textwidth]{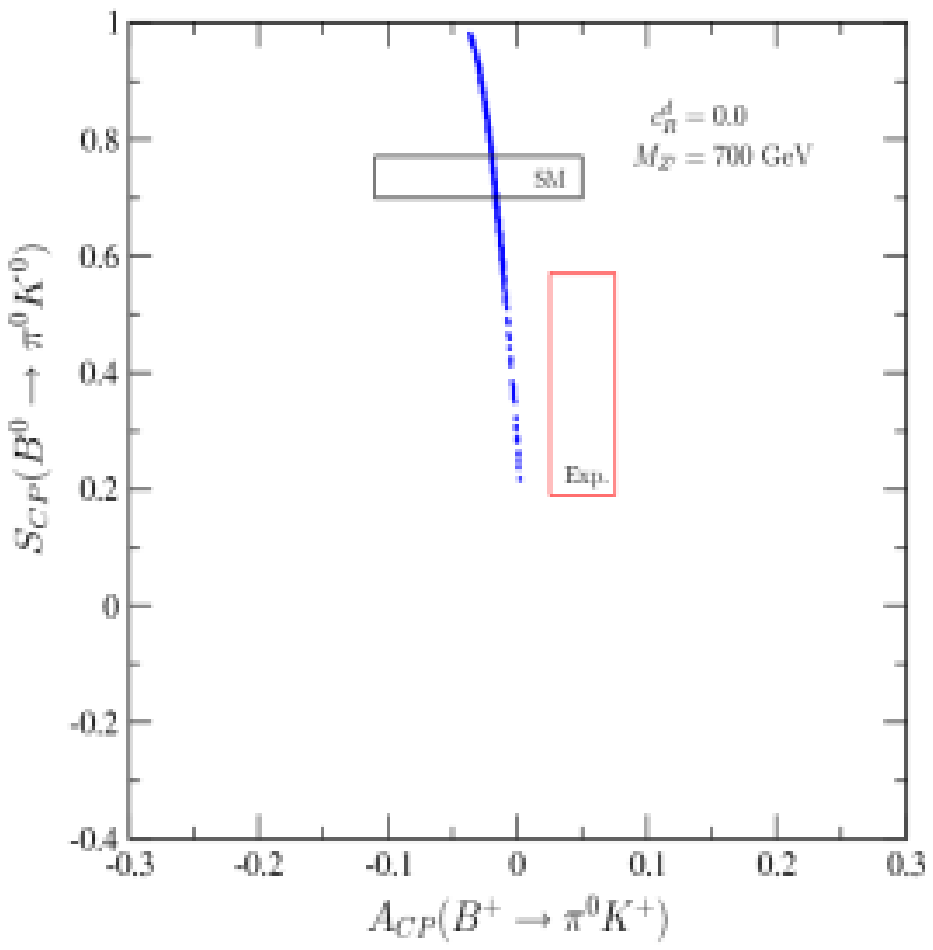}
}
\end{center}
\caption{
The correlation between
$A_{\rm CP}(B^+ \to \pi^0 K^+)$ and $S_{\rm CP}(B^0 \to \pi^0 K^0)$
for $M_{Z'}=700$ GeV
and (a) $c_R^d=1.0$,
(b) $c_R^d=0.5$,
(c) $c_R^d=0.0$.
}
\label{fig:SCP}
\end{figure}

The predictions for the correlation between
$A_{\rm CP}(B^+ \to \pi^0 K^+)$ and $S_{\rm CP}(B^0 \to \pi^0 K^0)$
are shown in Figure~\ref{fig:SCP}.
While it is difficult to get $S_{\rm CP}(B^0 \to \pi^0 K^0)$
as low as $\sim 0.38$ which is the central value for the current experiments,
it is possible to accommodate both $S_{\rm CP}(B^0 \to \pi^0 K^0)$
and $A_{\rm CP}(B^+ \to \pi^0 K^+)$ simultaneously
for $c_R^d=1.0, 0.5$ (Figure~\ref{fig:SCP}(a),(b)).
Again it is difficult to get large deviations from the
SM prediction for the $S_{\rm CP}(B^0 \to \pi^0 K^0)$ for $c_R^d=0.0$.
Therefore the value $c_R^d=0.0$, corresponding to
$P'_{\rm EW}(Z')=0$, is disfavored even if we include the
hadronic uncertainties in the calculation.

\begin{figure}
\begin{center}
\subfigure[]{
\includegraphics[width=0.25\textwidth]{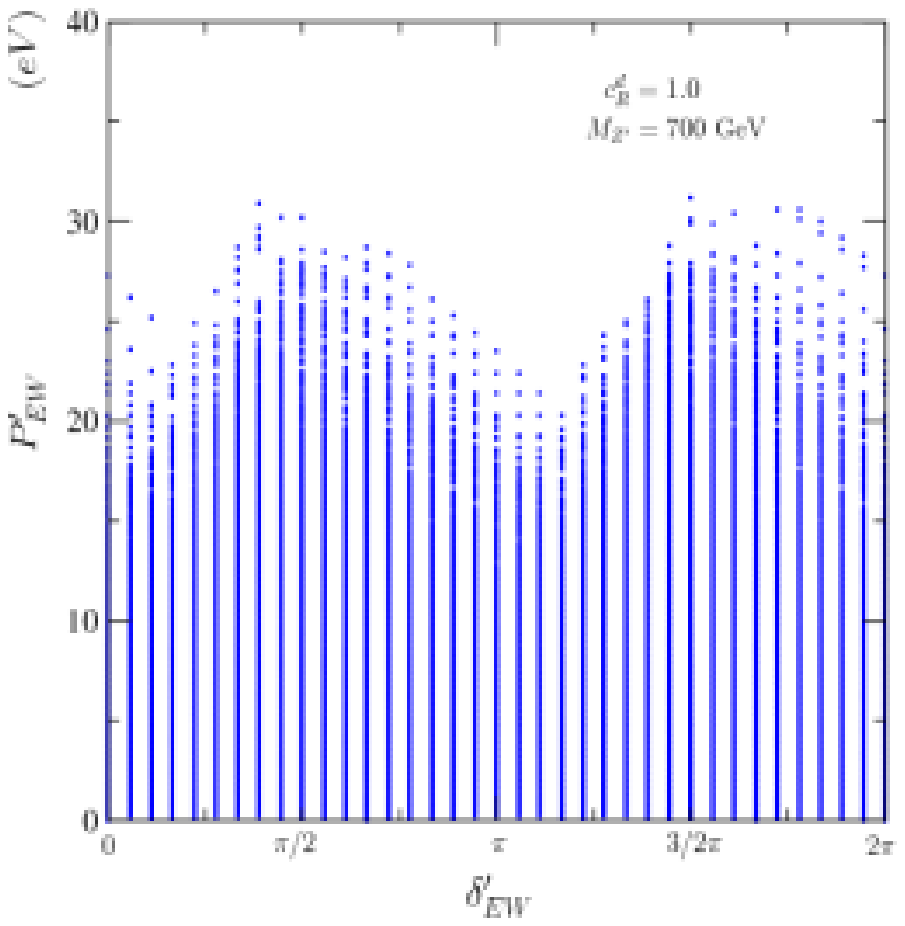}
}
\hspace{0.5cm}
\subfigure[]{
\includegraphics[width=0.25\textwidth]{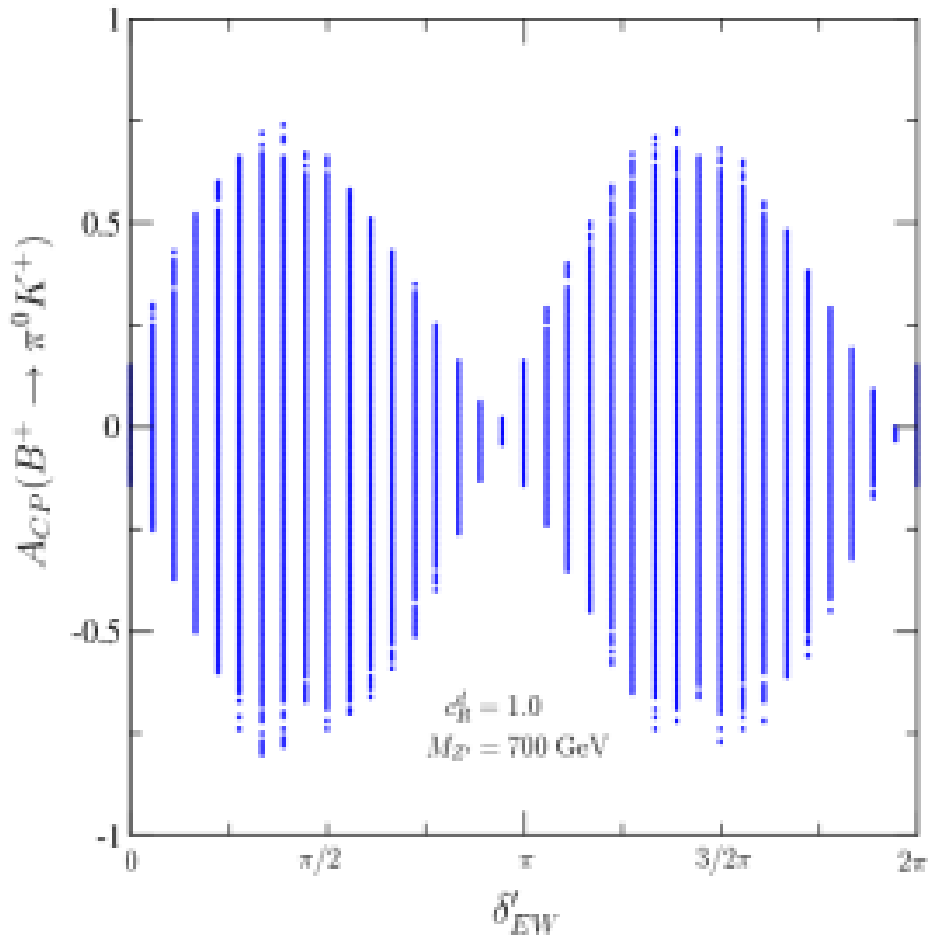}
}
\hspace{0.5cm}
\subfigure[]{
\includegraphics[width=0.25\textwidth]{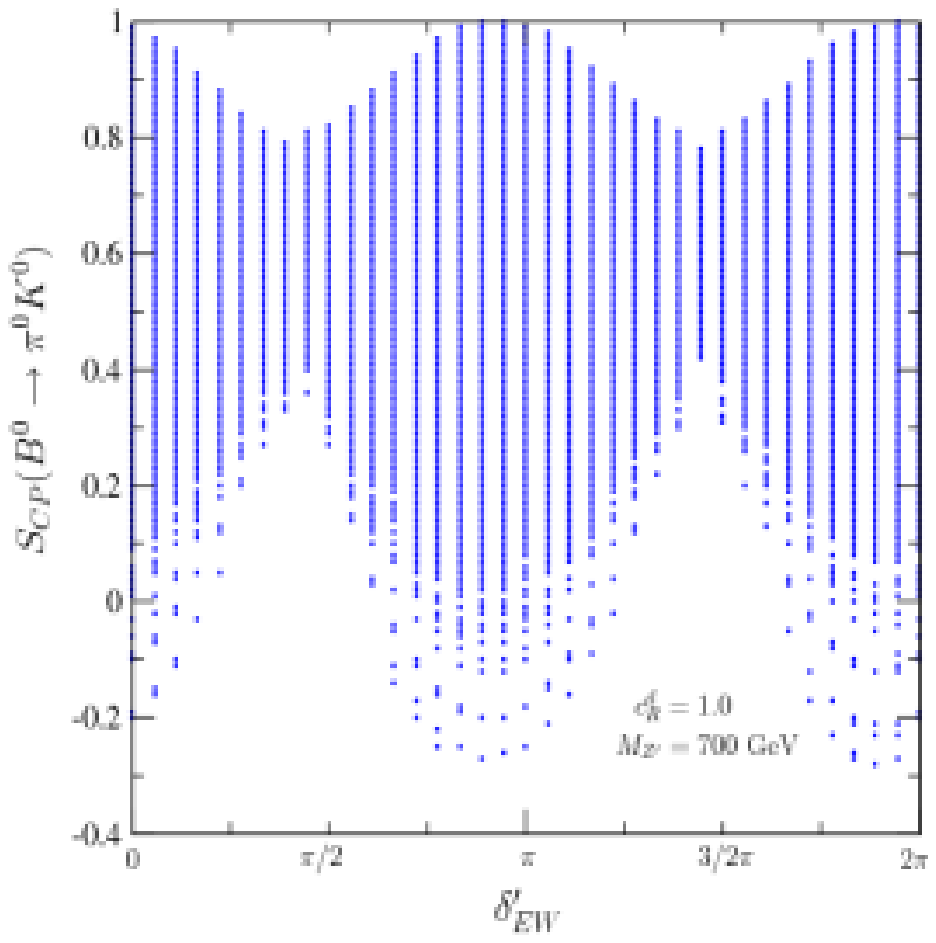}
}
\end{center}
\caption{
The $|P'_{\rm EW}|$ (a), $A_{\rm CP}(B^+ \to \pi^0 K^+)$ (b),
and $S_{\rm CP}(B^0 \to \pi^0 K^0)$ (c) as a function of
strong phase, $\delta'_{EW}$, of the electroweak penguin.
We fixed $c_R^d=1, M_{Z'}=700$ GeV.
}
\label{fig:delta}
\vspace{8pt}
\end{figure}

Until now we fixed the NP strong phases to be equal to the
corresponding SM strong phases. In general, they may not
be equal to each other. To see the effect of the NP
strong phases, now we allow the strong phase
of the dominant NP electroweak penguin $\delta'_{EW}$ to take
arbitrary values. Figure~\ref{fig:delta} shows that it can
give strong impact on the $A_{\rm CP}(B^+ \to \pi^0 K^+)$.
(For these plots we fixed $c_R^d=1, M_{Z'}=700$ GeV.)
The reason is that if $\delta'_{EW}$ is equal to the
strong phase of the dominant QCD penguin,
the NP electroweak penguin contribution to $A_{\rm CP}(B^+ \to \pi^0 K^+)$
vanishes at the leading order of $P'_{EW}(Z')/P'_{tc}$,
independent of the weak phases $\phi_{L(R)}^{Z'}$.
The effects of $\delta'_{EW}$ on $|P'_{EW}|$ and
$S_{\rm CP}(B^0 \to \pi^0 K^0)$ are rather minor,
and any value of $\delta'_{EW}$ can successfully
explain the $S_{\rm CP}(B^0 \to \pi^0 K^0)$ anomaly.

\begin{figure}
\begin{center}
\includegraphics[width=0.25\textwidth]{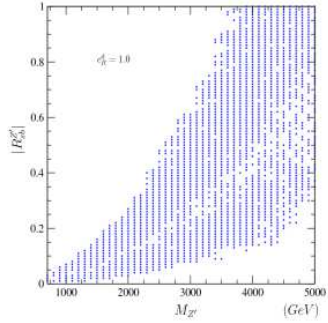}
\end{center}
\caption{ A scattered plot in ($M_{Z'}$,$|R^{Z'}_{sb}|$) plane. For
this plot we imposed $A_{\rm CP}(B^+ \to \pi^0 K^+)$, $A_{\rm
CP}(B^0 \to \pi^- K^+)$, and $S_{\rm CP}(B^0 \to \pi^0 K^0)$
constraints as well as the $\Delta m_s$ and $BR(B \to \pi K)$'s. }
\label{fig:MZprime}
\end{figure}
Now we consider the mass dependence of the $Z^\prime$ gauge
boson.  In Figure~\ref{fig:MZprime} we can see that the $M_{Z'}$ as large as
$5$ TeV which is beyond the LHC reach can accommodate the data with
$|R_{sb}^{Z'}| \lesssim 0.3$. (We fixed $c_R^d=1$ for the plot.)
The parabolic shape can be understood from (\ref{eq:Zprime-WC}) because
the parameter $|L_{sb}^{Z'}|$ can be fixed in terms of $|R_{sb}^{Z'}|$ by $\Delta m_s$.

\section{Conclusions}
\label{sec:conclusion}
We considered the leptophobic $Z^\prime$ model in the flipped
SU(5) GUT
obtained from heterotic string theory~\cite{LNY}.
This is phenomenologically interesting
because it contains ingredients which can possibly explain the apparent
deviations from the SM predictions in the $B\to \pi K$ decays:
\begin{itemize}
\item The new $Z^\prime$ coupling is generation dependent and can generate
FCNC processes.
\item The FCNC couplings allow large CP violation.
\item The couplings also violate the isospin symmetry and
can give large contributions to the electroweak penguins, $P'_{\rm EW}$
and $P^{'C}_{\rm EW}$.
\end{itemize}

We found that if we include the left- and right-handed FCNC couplings
$L_{sb}^{Z'}$ and $R_{sb}^{Z'}$ simultaneously, we cannot obtain the absolute
upper bounds for them contrary to~\cite{BJK_Zprime}
where it was assumed that only a single coupling exists at a time.
If we impose the additional
constraints, $BR(B \to \pi K)$'s, with some reasonable assumptions,
we can constrain $|L_{sb}^{Z'}|$ and $|R_{sb}^{Z'}|$.

We predicted the CP asymmetries, $A_{\rm CP}(B^+ \to \pi^0 K^+)$,
$A_{\rm CP}(B^0 \to \pi^- K^+)$, and $S_{\rm CP}(B^0 \to \pi^0 K^0)$.
Interestingly enough all of them are consistent with the current experimental
results when the isospin breaking coupling, $c_R^d$, is non-vanishing.
The case for $c_R^d=0$ where there is no $Z^\prime$ contribution to
the electroweak penguin is disfavored from the current data.

\vspace*{12pt}
\noindent
{\bf Acknowledgement}
The work of C.S.K. was supported  in part by  CHEP-SRC Program, and
in part by the Korea Research Foundation Grant funded by the Korean Government (MOEHRD)
No. KRF-2005-070-C00030.
The work was supported in part by the Korea Research Foundation Grant
funded by the Korean Government (MOEHRD) No. KRF-2007-359-C00009 (SB).
J.H.J was supported by the National Graduate Science $\&$ Technology Scholarship
funded by the Korean Government.

\noindent


\end{document}